# Scalable Estimation of Precision Maps in a MapReduce Framework

Claus Brenner
Institute of Cartography and Geoinformatics
Leibniz Universität Hannover
Appelstr. 9a, 30167 Hannover, Germany
claus.brenner@ikg.uni-hannover.de

## ABSTRACT
This paper presents a large-scale strip adjustment method for LiDAR mobile mapping data, yielding highly precise maps. It uses several concepts to achieve scalability. First, an efficient graph-based pre-segmentation is used, which directly operates on LiDAR scan strip data, rather than on point clouds. Second, observation equations are obtained from a dense matching, which is formulated in terms of an estimation of a latent map. As a result of this formulation, the number of observation equations is not quadratic, but rather linear in the number of scan strips. Third, the dynamic Bayes network, which results from all observation and condition equations, is partitioned into two sub-networks. Consequently, the estimation matrices for all position and orientation corrections are linear instead of quadratic in the number of unknowns and can be solved very efficiently using an alternating least squares approach.

It is shown how this approach can be mapped to a standard key/value MapReduce implementation, where each of the processing nodes operates independently on small chunks of data, leading to essentially linear scalability. Results are demonstrated for a dataset of one billion measured LiDAR points and 278,000 unknowns, leading to maps with a precision of a few millimeters.

## 1. INTRODUCTION
Since more than 20 years, digital maps have been used to support car and personal navigation systems. Recently, the development of highly detailed and accurate maps has gained momentum, since such maps are required for advanced driver assistance systems, as well as partially or fully autonomous cars. In order to collect such maps, mobile mapping systems are employed, which usually combine vision sensors, such as cameras and laser scanners (LiDAR) with a localization subsystem. In case of vehicles, the latter is normally a combination of a global navigation satellite system (GNSS) receiver, an inertial measurement unit (IMU) and an odometer (wheel distance measurement). All measurements are combined by a filter approach to deliver a continuous position and orientation (pose) update, at a typical rate of 200 Hz.

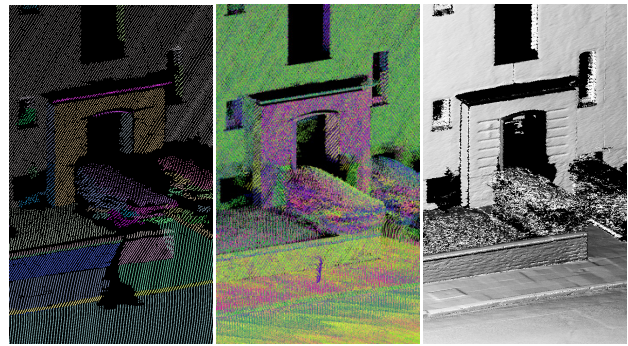

**Figure 1. Example results.** A large number of single scans (strips) acquired by LiDAR mobile mapping (left) are aligned (middle) using a global optimization. After alignment, systematic errors are removed which allows estimating the surface with a high density and a standard deviation of only a few millimeters. Details such as façade and wall structure and sidewalk pavement become visible (right).

Even if highly accurate (and expensive) GNSS/IMU components are employed, and measurements are processed using differential GNSS post-processing, one can typically observe absolute errors in the range of 30 centimeters in urban areas. While this seems to be fairly accurate, it is in large contrast to the accuracy of the LiDAR sensors themselves, which nowadays have down to 5 mm accuracy (and 3 mm precision), a difference of two orders of magnitude. Since contemporary LiDAR mobile mapping systems reach measurement rates of up to 2 million measured points per second, which is 10,000 observations for each pose delivered by the (200 Hz) GNSS/IMU system, it is highly attractive to correct the pose using the LiDAR measurements themselves.

## 2. RELATED WORK
The problem of adjusting sensor poses using measurements is fundamental to geodesy and surveying. In photogrammetry (and lately, computer vision), it is known as bundle adjustment, and after several decades of research, it is still the dominant refinement approach, due to its rigorous formulation of the functional and stochastic error model [19]. In robotics, such problems are usually encountered in the context of simultaneous localization and mapping (SLAM, [7]), where (in-) dependencies between random variables are typically depicted using probabilistic graphical models [12], especially (dynamic) Bayes networks. It is important to note that the frequent problem of iteratively estimating the 'current best state', also known as *on-line SLAM*, is different from the so-called *full SLAM* problem, where *all* unknowns are to be estimated simultaneously, using all observations. Since the main objective of this paper is to improve maps, rather than to find the current best state of the mobile mapping system, the problem is of *full SLAM* type.

A special estimation problem that has received large attention is the alignment of data, especially the alignment of 2D surfaces in 3D space. Since no 3D point correspondences are known in this case, tentative correspondences are chosen, which are used to estimate the optimal transformation, which in turn is used to establish new point correspondences. This is known as the *iterative closest point* (ICP) algorithm [2], and different strategies have been devised to select suitable correspondences [17]. Apart from the strategy how corresponding pairs are selected, it is also important to decide which functional model is introduced. In its basic form, the ICP algorithm uses 3D point correspondences to estimate a rigid or similarity transformation, i.e., all three components of the residual vector are used during minimization. However, since 2D surfaces are aligned, which are locally planar, only one component should be introduced into the functional model, namely, the distance parallel to the surface normal. A similar effect can be obtained by weighting of the quadratic error term, known as generalized ICP [18]. Its advantage is that it is a direct extension of the standard ICP algorithm which requires only minor modification. However, if the introduction of variances in the error term is only used to achieve the 'point-to-plane effect', this leads to computational overhead. Therefore, in the model presented below, a direct formulation of the point-to-plane observation equation is preferred.

In photogrammetry, as well as rigid body alignment, the basic 'building blocks' are assumed to be rigid structures (the bundle of rays, and the surface, respectively). However, the typical characteristic of dynamic LiDAR mobile mapping is that the sensor system undergoes continuous movement while scanning. This leads to non-rigid ICP formulations. In case of LiDAR, this is also termed *strip adjustment* and has been investigated since the advent of airborne LiDAR systems, 20 years ago [9]. It is still a contemporary research topic in the context of unmanned aerial vehicles with LiDAR sensors [6]. Similar problems occur with rolling shutter cameras on moving platforms. One solution that has been applied at a global scale is to consider the IMU trajectory as being 'correct' for short time intervals and to 'bake' it into each exposure span (one full 'roll' of the shutter) [11]. However, for a rigorous error modeling, strip adjustment is often formulated by estimating pose correction parameters at constant time or distance intervals along the trajectory, constraining the relative pose changes by condition equations. This leads to a large number of unknowns, which in turn result in large, sparsely populated estimation matrices.

While in traditional airborne laser scanning, the number of overlapping scan strips may be low, it is more common in mobile mapping to have a large number of overlapping strips, acquired during different measurement campaigns. For $n$ overlapping strips, $O(n^2)$ pairings are possible, each of which leads to a corresponding observation equation. A standard solution to this problem is to introduce additional unknowns, *tie points*, so that instead of relating every strip to every other strip (leading to $O(n^2)$ equations), every strip is related to the tie point (leading to only $O(n)$ equations). Since actually, surface elements need to be matched, rather than individual points, small *tie surface patches* need to be introduced as additional unknowns. This approach has been described by Huang and Anguelov [8], who termed the collection of surface elements, in Bayesian parlance, the *latent map*.

Different approaches are possible to represent the latent map. In [8], the scene is subdivided into small grid cells, each of which represents one line. With this representation, it was observed that thin surfaces, captured from both sides, lead to wrong assignments. Instead of raising the density of the grid, which was found to decrease computational efficiency drastically, the authors opted for a representation with a maximum of two lines in each grid cell, which are distinguished by their opposite normal vectors. Recently, implicit surface representations in the form of truncated signed distance functions (TSDF), originally proposed by Curless and Levoy [3], have gained increased attention. Newcombe et al. [13] have shown how TSDFs can be embedded into an incremental mapping process, using the Microsoft Kinect sensor. In each step, the current surface reconstruction is used to estimate the pose of the sensor; subsequently, the sensor measurements are used to improve the surface estimate by updating the TSDF. Being incremental, this method does not perform a global optimization. Since the TSDF represents the surface implicitly by a scalar function $f(x, y, z) = 0$, represented by a discrete scalar field, its memory requirements are cubic in the scene extents. However, since the distance function is truncated, it will be undefined for most of the space. Therefore, spatial data structures have been used to efficiently store the TSDF only in areas where it is defined. Recent approaches favor voxel hashing, where small sub-blocks of $8^3$ or $16^3$ scalar values are hash-indexed, leading to much lower memory requirements while still having $O(1)$ access time [14][10].

## 3. MOBILE MAPPING DATA

The test area is located in Hannover, Germany. Over time, it was covered by several mobile mapping measurement campaigns. From these, scan strips around a central 'double-8-loop' were selected (see Figure 2). The data was not cut at the borders of this area, so that strips extend beyond the 'double-8'. Overall, 150 scan strips were selected (75 pairs). The number of scan strips which cover an individual street varies between two and 28.

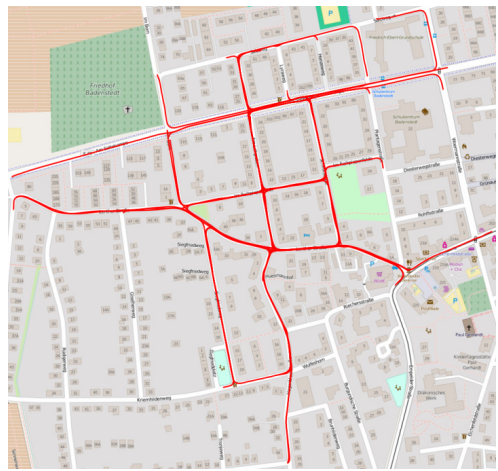

**Figure 2. Test area in Hannover, Germany (scan trajectories marked in red). Scan strips were selected around a central 'double-8' loop. Background: OpenStreetMap.**

The data was acquired using a RIEGL VMX-250 mobile mapping system [16]. This system includes two 360° LiDAR scanners, each capable of measuring 300,000 points per second (pulse rate) at 100 profiles per second (rotation rate), with a specified accuracy of 10 mm and precision of 5 mm (both 1σ). For positioning, an Applanix POS-LV 510 system is used, which integrates a GNSS receiver, IMU, and odometer. It is specified with 2 cm accuracy in position and 5 cm in height, and angular errors of 0.005° (roll and pitch) and 0.015° (true heading), where all values are RMS, and the data is post-processed using GNSS reference stations. For GNSS outages of 60 seconds, the system is specified with 10 cm error in position and 7 cm in height [1]. In practical situations, for urban areas with buildings and trees of moderate height (as in the case of the test area used here), one can expect position and height errors to reach 20-30 centimeters.

## 4. SCAN STRIP SEGMENTATION

Instead of converting the scan strip data immediately to 3D point clouds, a segmentation into continuous regions is performed first. The basic rationale is that the point clouds are acquired in a highly structured fashion, using rotating mirrors and forward motion of the vehicle. Therefore, the acquisition sequence defines a topology between the scanned points, which would get lost if the data were converted into unstructured 3D point clouds.

In order to segment the strips, the data is put into a regular raster of *pixels:* 3000 lines (the number of measurements per scan head revolution) times the number of columns which depends on the duration of the acquisition (for each second, 100 columns). For each pixel, a normal vector is estimated. A robust, random-sampling consensus (RANSAC) based plane estimation is used, which preserves sharp edges. The point and normal vector data is then used to compute a homogeneity criterion, which evaluates $C^0$ and $C^1$ continuity. Then, the efficient, graph-based image segmentation approach of Felzenszwalb and Huttenlocher [5] is used to obtain image regions. Note that while the method uses the 'image topology' of the points in the scan strip, the homogeneity criterion is solely based on the underlying geometry.

Figure 3 and Figure 4 show typical results for one scan strip (with colors chosen randomly). As one can see, the algorithm manages to find the major structures, such as the street and sidewalk surfaces, but also minor details such as façade parts and curbstones. Note that while the homogeneity criterion evaluates continuity, it is not a segmentation into planar patches, which would partition the curved road surface into many regions.

As a net result of this pre-processing, connected regions are available which form surfaces in 3D space, with a normal vector defined at each surface point. Small area regions are removed, which also reduces tree foliage to a large extend, and helps to avoid wrong assignments in the later matching steps.

## 5. ESTIMATION
### 5.1 Estimation Model

A typical situation of the data capture situation is depicted in Figure 5. The blue scan vehicle travels along the dark blue trajectory. The red lines indicate the scan planes of the two LiDAR scanners. The scan ray hits an object, or *map element m*, but due to errors in the measurement of the vehicle trajectory, it does not indicate the correct position. Similarly, at a different time instance, the green vehicle hits the same map element. The green and blue trajectories may also be obtained during the same scan, when *m* is hit first by scanner 1, and later by scanner 2. The task is to align the two scanned points at *m* by deforming the green and blue trajectories, earlier introduced as strip adjustment. This deformation is accomplished by estimating correction terms to the pose at discrete locations $a_1, a_2, a_3, ...$ and $b_1, b_2, b_3, ...$, called *anchor points*. The spacing of anchor points should be selected according to the expected characteristics of the data, either in space or time. In this paper, a spatial distance of 0.5 m was chosen, which means that for any part of a trajectory, for which LiDAR observations are available that lead to minimization terms, six unknowns are introduced every 0.5 m along the trajectory. Between anchor points, the pose parameter corrections are interpolated.

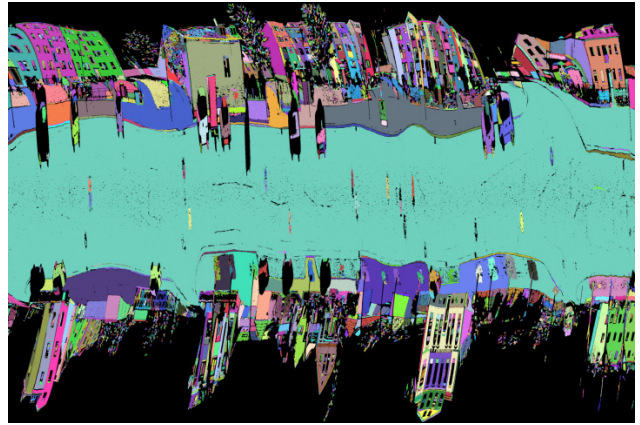

**Figure 3. Example for a segmented scan strip. Segments were assigned a random color.**

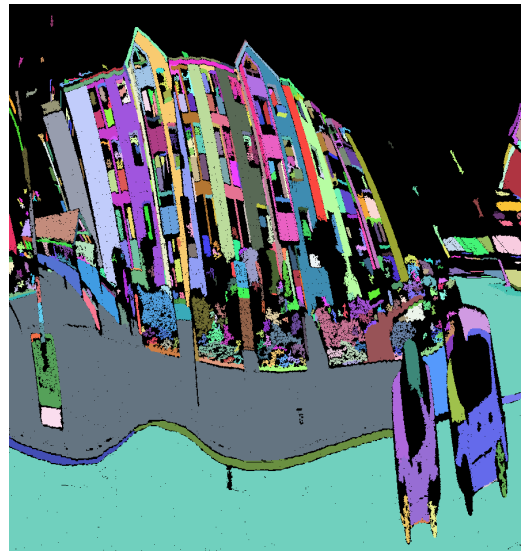

**Figure 4. Detail of Fig. 3. Façade structures, walls, car parts and curbstones can be discerned. Note that small regions are removed.**

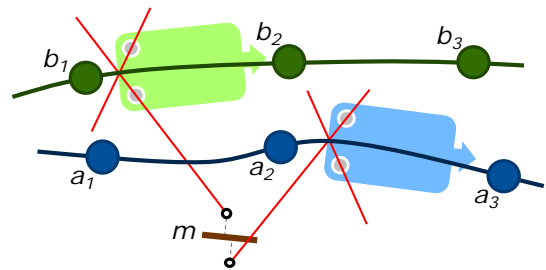

**Figure 5. Illustration of a scan situation.**

Since the surface element $m$ is not known, one would typically minimize the distance between the two scan points. As noted earlier, if $n$ scans are available, this would lead to $O(n^2)$ observation equations. Instead, the surface element $m$ is introduced as an additional unknown, so that only $O(n)$ observation equations are needed to relate the scan points to $m$.

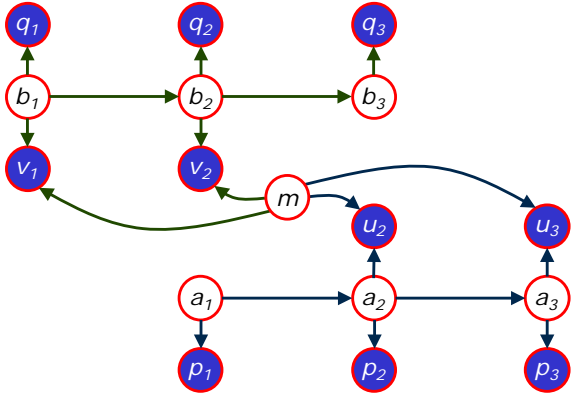

**Figure 6. Bayes network for the situation shown in Figure 5.**

To illustrate the overall situation, Figure 6 shows the Bayes network for the scene in Figure 5. In this case, one trajectory $a$ is related to another trajectory $b$ by a single map element $m$. The task is to estimate

$$\hat{x} = \underset{a_1,a_2,a_3,b_1,b_2,b_3,m}{\mathrm{argmax}} P(a_1, a_2, a_3, b_1, b_2, b_3, m) \quad (1)$$

where $\hat{x}$ denotes the estimate of the vector of all random variables $a_1, a_2, a_3, b_1, b_2, b_3, m$. For the purposes of this paper, all random variables are assumed to be normal distributed and the solution is found using least squares adjustment.

The poses in $a_i$ were observed by the GNSS/IMU system, which is introduced as observed variables, $p_i$ (similarly for $b_i$, which are observed by $q_i$). Since corrections are estimated rather than absolute poses, this leads to observation equations of the form

$$0 + v = a_i \;\; ; \Sigma_{\mathrm{prior}} \quad (2)$$

where $\Sigma_{\mathrm{prior}}$ is the variance of the GNSS/IMU measurement and $v$ is a placeholder for the residual which is to be minimized. Note that these are actually six observation equations (three for position, three for rotation angles). Similar equations are required for all $b_i$. In addition, successive anchor points are constrained to have similar pose corrections. This enforces a smoothing of the pose correction along the trajectory and leads to equations of the form

$$0 + v = a_{i+1} - a_i \;\; ; \Sigma_{\mathrm{smooth}} \quad (3)$$

for all anchor elements which are actually used (similarly for $b_i$). $\Sigma_{\mathrm{smooth}}$ can be set according to the assumed drift of the trajectory. In practice, it is usually used as a design parameter to control the trajectory smoothness. Again, these are six equations for any anchor pair which takes part in the estimation. Finally, the observed random variables $u_i$ relate the pose corrections $a_i$ to the map element $m$, which is also unknown (similarly for $v_i$, which relate $b_i$ to $m$). As described earlier, since the location is not constrained along the surface, but only perpendicular to it, only a scalar observation equation is necessary,

$$0 + v = \langle w_j, s_j - (R_i r_k + t_i + t_{0,k}) \rangle \; ; \sigma_{\mathrm{dist}}^2$$

which, bringing constant terms to the left hand side, leads to

$$\langle w_j, s_j - t_{0,k} \rangle - v = \langle w_j, R_i r_k + t_i \rangle \; ; \sigma_{\mathrm{dist}}^2 \quad (4)$$

where $w_j$ is the surface normal vector and $s_j$ is a point on the map element $m_j$, $r_k$ is the measurement vector from the LiDAR scan head to the measured point and $t_{0,k}$ is the associated scan head position, so that without correction, $r_k + t_{0,k}$ would be the measured point. This is corrected by the translation $t_i$ and rotation $R_i$, which are part of the pose correction $a_i$. Since $t_i$ and $R_i$ are interpolated from the anchor points, this equation is set up for $a_i$ and $a_{i+1}$, if the measurement took place while the vehicle was between anchors $a_i$ and $a_{i+1}$. The variance $\sigma_{\mathrm{dist}}^2$ is set according to the distance measurement accuracy of the LiDAR scanner.

## 5.2 Estimation procedure

The overall estimation proceeds iteratively. It uses the estimated trajectory corrections from the previous pass to correct the entire point cloud. Then, map elements are defined which are used to set up all correspondences, which in turn lead to estimation equations for the distances (Eq. 4). These are complemented by the constraint equations for the prior (Eq. 2) and smoothness (Eq. 3). If all equations, written down in rows, are denoted by the design matrix $A$, the solution can be computed by solving the normal equation system

$$\hat{x} = (A^T P A)^{-1} A^T P l, \quad (5)$$

where $P$ is the weight matrix of the observations which results from all variances given in Eqs. 2-4, and $l$ is the column vector of all left hand sides of these equations (which is nonzero only for Eq. 4). Note that the actual procedure to obtain $\hat{x}$ is different, as discussed in the sequel.

Since the matching is dense, the number of rows of $A$ is the number of LiDAR observations (Eq. 4), plus the number of constraint equations (Eqs. 2 and 3), which is linear in the distance travelled. Of those, the number of LiDAR observations is by far dominant. For example, when the mobile mapping van travels at an average speed of 10 m/s (36 km/h, 22 mph), 20 anchor points are generated per second (based on one anchor every 0.5 m distance), which leads to a total of 120 constraint equations, as opposed to a maximum of 600,000 measured LiDAR points. Even if only 30-50% of the LiDAR measurements lead to distance equations, there are about three orders of magnitude more equations of type (4) than of type (2) and (3).

In the example presented here, this amounts to approximately one billion rows in $A$. Though this looks daunting, it is not a limiting factor, since the standard approach to set up Equation 5 is to directly compute the sub-blocks for $(A^T P A)_{\mathrm{block}}$ ($6 \times 6$ matrices) and $(A^T P l)_{\mathrm{block}}$ (6-vectors) upon processing one observation equation, which are then added up in an overall matrix $A^T P A$ (which is sparse), and vector $A^T P l$, both of which have as dimension the number of unknowns, rather than observations.

Concerning the unknowns, these consist of the pose correction terms for each anchor, but also of the map elements $m$, as can be seen in Figure 6 and Eq. 1. This raises the question how the map elements are represented. In Eq. 4, the only relevant properties of a map element, in terms of the observation equation, are the normal vector $w_j$, and the distance of the transformed LiDAR point to the surface. A suitable and conceptually simple surface representation would be in the form of a TSDF. However, even if the TSDF is made sparse by using relatively small blocks, it is inherently a volumetric representation. For standard scenes, which contain large areas of flat surfaces, this wastes a lot of memory.

Therefore, an approach based on local surface models was used. Similarly to a TSDF, the space is subdivided into cells of equal size, which are indexed using voxel hashing. However, inside each cell, no 3D grid of distance and weight values is used, but rather, the surface is represented by a set of local surface models (LSM), each modeling a height field over a regular 2D raster in terms of 'height pixels'. The number of LSMs inside each cell is determined by the number of different surface normal vectors.

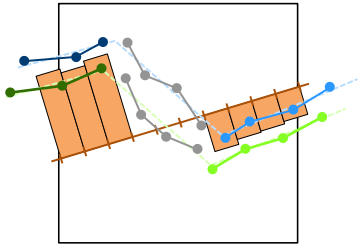

**Figure 7. Representation of one local surface model, part of a map element. Illustration in 2D, actual representation is a 3D cell with a height model over a 2D raster.**

Figure 7 illustrates this representation. Imagine the real situation is a 'curbstone' so that the left part is the sidewalk, the right part is the road surface, and the (steep) middle part is the shoulder of the curbstone. Two LSMs will be initialized in this case, one for the road and sidewalk surface, and another one for the shoulder of the curbstone (not shown here). All blue points stem from one scan. Since their connectivity is known from the pre-segmentation (section 4), the affected LSM pixels can immediately be identified; no further triangulation is necessary. Moreover, since the segmentation separates the different regions, wrong assignments are minimized. For example, the grey points belonging to the curbstone shoulder do not affect the LSM for the sidewalk/ road surface. Every scan (e.g., also the green one in Figure 7) influences the associated LSM. In summary, the global map consists of map elements, which are 3D cells, each of which contains one or more LSMs which represent the surface in terms of a 2D raster of height values.

Computing this map amounts to the estimation of all height values (orange bars in Figure 7) in all LSM cells. Since the LSMs use a 2D grid of constant spacing, the overall number of unknowns is proportional to the total surface of the scene. Again, this is the dominant term. In the example presented here, there are 278,052 unknowns due to the anchor points, but millions to billions of unknowns due to the map elements (depending on LSM grid resolution). Such a number of unknowns may still be handled by a sparse solver, however, as this grows linearly with the size of the scene, handling and solution of the normal equations would have to be spread across multiple machines. The following section describes how this can be realized using the MapReduce framework [4] instead.

## 6. SCALABLE ESTIMATION
### 6.1 Partitioning the Dependency Graph

The Bayes network shown in Figure 6 allows to reason about the dependencies between random variables. For example, if $a_3$ changes, this will influence $m$, since $u_3$ is observed and arrows meet head-to-head. By a similar argument, this will influence $b_1$ and so on, so that overall, the change spreads across the entire graph.

The key observation is that if $m$ is observed, this chain of dependencies is broken, so that a change in $a_3$ only influences the anchors $a_2$ and $a_1$. Conversely, if all $a_i$ and $b_i$ are observed, the map elements $m$ become independent. Therefore, similar to the approach in [8], the unknowns are split into two groups. Instead of maximizing

$$X^* = \underset{A,M}{\operatorname{argmax}} P(A, M)$$

where $A$ and $M$ are the anchor point and map unknowns, respectively, the two maximizations

$$A^* = \underset{A}{\operatorname{argmax}} P(A|M) \qquad (6)$$

$$M^* = \underset{M}{\operatorname{argmax}} P(M|A). \qquad (7)$$

are iterated. In the context of the models used here, this is also known as the *alternate least squares* approach.

Concerning Eq. 6, as just demonstrated for the example in Figure 6, if the map $M$ is held fixed, the remaining dependencies are only between anchors of the same trajectory. As this graph is linear, it can be solved exactly and efficiently by a single forward-backward pass of a max product algorithm [12]. In case of continuous, normal distributed variables, this algorithm is known as *Kalman smoother*, or *Rauch-Tung-Striebel* algorithm [15]. It consists of a forward pass, identical to the Kalman filter, and an additional backward pass. Note that this is not an approximation, but the exact maximum likelihood estimate for the unknowns $A$.

For the map optimization, Eq. 7, the situation is even simpler. Since all map elements are independent, they can be estimated individually. This is done using a least squares estimation, yielding a mean value and variance for every pixel of every LSM (every orange bar in Figure 7).

### 6.2 MapReduce: Pre-Processing

The overall process is mapped to two separate MapReduce calls, the pre-processing, called once, and the estimation, which is iterated. Pre-processing is shown in Figure 8. It is assumed that raw data is available in the form of individual scan strips. In this work, scan strips are bounded in length, due to the operation of the mobile mapping system. Otherwise, they could be cut at arbitrary positions into records of manageable size. The task of one mapper is to compute first the segmentation (section 4). Since each individual strip is bounded in size, this can be done in memory. The time complexity of Felzenszwalb and Huttenlocher's algorithm is $O(n \log n)$, where $n$ is the size of the individual scan strip [5]. It can be argued that, since the strip size is bounded, the processing of an individual strip is of constant time complexity, $O(1)$. Therefore, the overall time complexity, in terms of the (more important) number of strips $N$, is linear, $O(N)$.

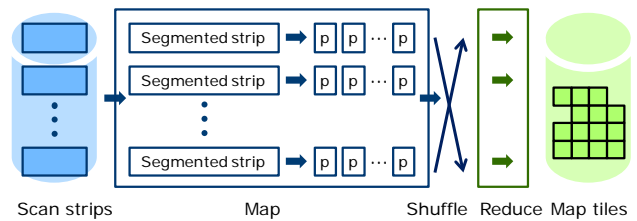

**Figure 8. MapReduce for the preprocessing step.**

After the segmentation is computed, the mapper assigns each single point a key which is the index of its spatial tile. This key is trivially computable from the global point coordinates and a given tile size. In the example shown here, this was selected to be 15 m. Since during the later process of estimation, the original mapping trajectories are deformed by the estimated pose corrections, the assignment of LiDAR points to tiles may be not correct anymore. To solve this, the mappers emit points along the borders to multiple tiles. Thus, the later process of estimation can proceed using the tiles only and does not need to go back to the original scan strips. Note that during estimation, these multiply emitted points do not pose a problem since they are ultimately used only in one tile. The border area should be selected according to the maximum shift expected. In the presented example, an overlap of 0.3 m was used, which, for a tile size of 15 m, corresponds to 8% overhead.

On the basis of this key, shuffling will send all points to their appropriate tile so that there is no work left for the reducers apart from emitting them. Note that the points in each tile consist not only of their X, Y, Z co-ordinates, but also contain e. g. the point along the trajectory when the measurement was taken, the segment id they belong to, and their index in the scan strip image, which allows recovering their neighbors in the original scan strip image.

### 6.3 MapReduce: Estimation

During estimation (shown in Figure 9), the set of tiles is the input to the mapping phase. In addition, the trajectory corrections from the previous iteration are required. The latter is relatively little data and can be efficiently broadcast.

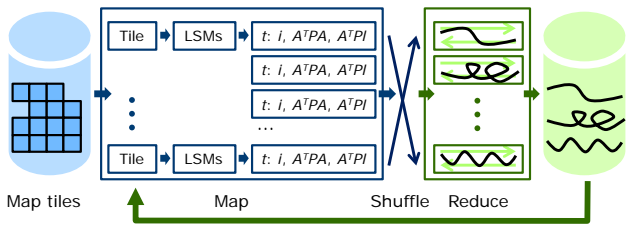

**Figure 9. MapReduce for the estimation step. Tiles lead to collections of LSMs, which lead to $(A^TPA)_{block}$ and $(A^TPl)_{block}$ blocks with key $t$ (the trajectory) and index $i$ (the anchor in the trajectory).**

While a single mapper processes one tile, it builds up the set of occupied cells, and sets up and updates the LSMs in each cell. This estimates the map, i.e., implements Eq. 7. Subsequently, for each measurement in the cell, the distance and normal vector can be obtained, required for the distance equation (Eq. 4). This generates the $(A^TPA)_{block}$ and $(A^TPl)_{block}$ blocks, which are valid for a certain anchor point of a certain trajectory. The mapper emits those blocks with the key being the trajectory id.

After the shuffle phase, every reducer will get all $(A^TPA)_{block}$ and $(A^TPl)_{block}$ blocks for a certain trajectory and will add them up accordingly. Since, as observed above, trajectories are independent when the map is given, each reducer is able to compute one part of the maximum likelihood solution (Eq. 6) independently from all others. The Rauch-Tung-Striebel algorithm is run and the pose corrections for each anchor point are obtained. These are then broadcast and the next iteration commences.

In terms of memory consumption, each mapper sets up the latent map for one tile only. After the observation blocks for one tile are emitted, it is discarded. The required amount of memory can be controlled by varying the (spatial) tile size. The achievable parallelism is equal to the number of tiles, which is ideal.

After shuffling, each reducer integrates incoming observation blocks into a sorted data structure, e.g. a balanced tree. This structure takes space linear in the length of the trajectory. During the forward pass of the Rauch-Tung-Striebel algorithm, matrices are generated which are updated during the backward pass. This takes also linear memory space. Note that during the entire algorithm, no large and/or sparse matrices are set up. In consequence, the estimation is linear in time and space, relative to the length of the mapping trajectories. The achievable parallelism is equal to the number of trajectories, which, again, is ideal. (Note that, as described in section 6.2, the length of the strips, and thus of the trajectories, is bounded, so that large projects will have thousands of trajectories.) In general, solving Eq. 6 is so fast that it is not relevant for the overall time budget.

## 7. RESULTS
### 7.1 Quantitative Results

The algorithm was applied to the dataset described in section 3. The pre-processing reads 150 scan strips and generates 1,287 map tiles of size $15 \times 15\ m^2$. The number of points per tile varies from one to 17.6 million (M). Overall, including the data introduced due to the tile overlap, there are 1.05 billion points in the dataset, which corresponds to 60 GB of raw data.

During estimation, 18 iterations were carried out. The adjustment is robustified using outlier removal. A distance threshold is introduced, which discards correspondences of points to latent map elements. Initially, this is set at 0.3 m and is reduced during iterations, down to 7 mm in the final iteration. As a result of this, the number of points which take part in the estimation varies. It starts at 898 M, then, due to alignment, increases to 923 M, from which it decreases, due to the increasingly tight threshold, to 849 M in the final iteration. For a more efficient computation, the grid size of the LSM cells is also varied, from 10 cm in the beginning down to 1 cm in the final adjustment, with most iterations using 2 cm. At 2 cm grid size, the reference implementation (C++), run on an Intel i7-4790k (4 cores), takes approximately 370 seconds per iteration (including reading of the 60 GB of raw data).

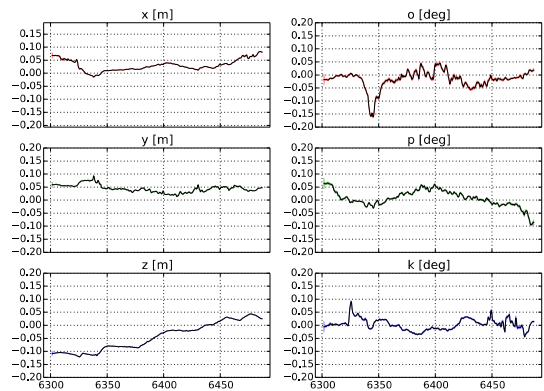

**Figure 10. Example for the estimated pose correction along one trajectory, between kilometer 6.3 and 6.5. Left column: shift, right column: rotation angles around omega, phi and kappa (heading).**

During each iteration, corrections for all trajectories are estimated. Overall, there are 75 trajectories, Figure 10 shows one example. Typical position corrections are within 0.2 m and angle corrections are within 0.2 degrees.

In order to assess the goodness of the fit, the standard deviation is computed in each latent map element. Figure 11 shows the latent map for the entire scene (c.f. Figure 2), before estimation. Standard deviations of 7 mm and above are shown in red, which leads to a mainly red plot. Some areas are green or blue, which happens mostly when some parts of the scene were mapped only once, so that they are covered only by two scan strips from the same trajectory, which are usually well aligned.

Figure 12 shows the same plot after the last iteration, using the same color scale. As can be seen, most areas are green, which corresponds to a standard deviation of approximately 3 mm.

outages are more probable. The latter is relatively wide with only a few, isolated trees.

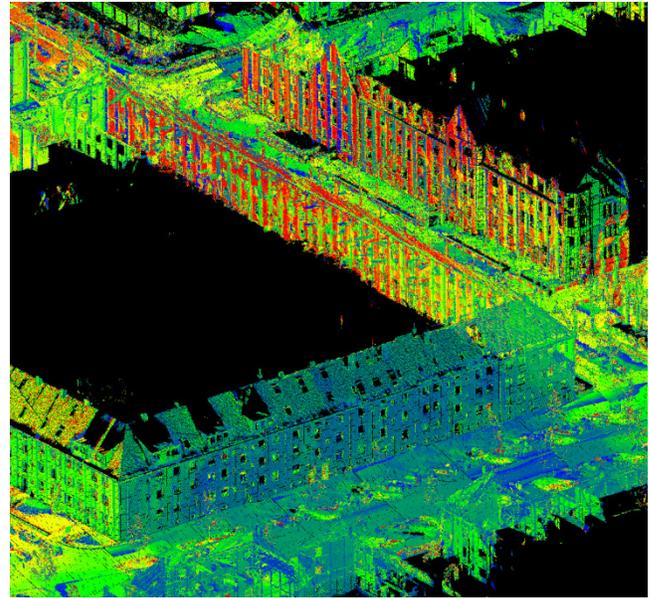

**Figure 13. Tilted view of a part of the scene, before estimation. Color scale: standard deviation, temperature scale, blue (0 mm) to red (100 mm or more).**

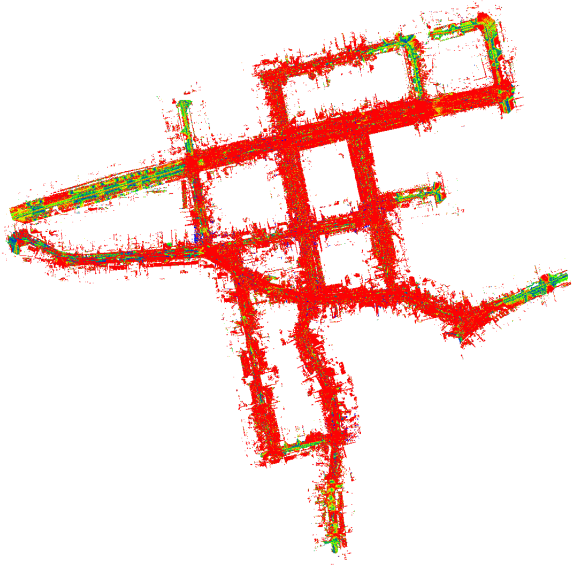

**Figure 11. Latent map of the project area, before estimation (top view). Color scale: standard deviation, temperature scale, blue (0 mm) to red (7 mm or more).**

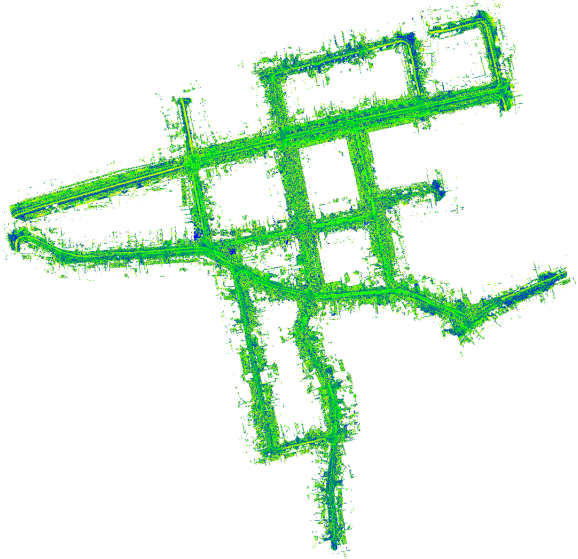

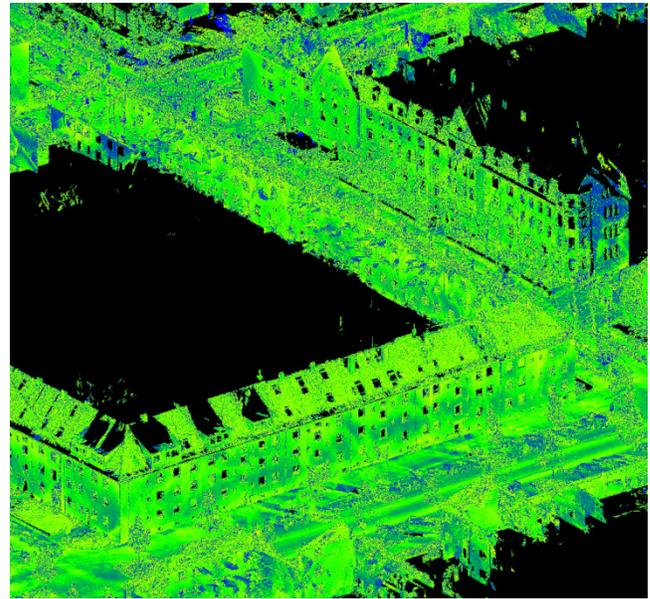

**Figure 14. Same view as Figure 13, after estimation. Note the color scale has been changed to the range between 0 mm (blue) to 7 mm (red).**

Figures 15 and 16 show a comparison for a close-up view. Note that in Figure 16, small details can be seen. Especially for the building in the lower left corner, a façade structure becomes apparent, which is due to the wall being covered by shingles. Compare this to Figure 20.

For a quantitative assessment, the distance between all LiDAR points and their corresponding latent map element was computed. Figure 17 shows histograms of the signed distance for some of the adjustment iterations. Since approx. 900 M points take part, the curves are very smooth. The bucket size is 0.1 mm, so that the

**Figure 12. Latent map of the project area used in the last estimation (top view). Color scale: standard deviation, temperature scale, blue (0 mm) to red (7 mm or more).**

Figures 13 and 14 show a tilted view of a part of the scene. It can be seen that before the iteration (Figure 13), there are large errors (0.1 m and more) present, while after estimation (Figure 14), the standard deviations are mostly reduced to a few millimeters. Note that before estimation, the street in the upper right corner shows larger errors than the street in the lower part of the figure. This coincides with the expectation, since the former is a narrow street with an alley of large trees and dense tree canopy, so that GNSS value of the highest (yellow) peak means that 14.6 M points were within 0 and +0.1 mm distance to the surface (the same holds for the -0.1 to 0 mm bucket). As can be seen, the distances are initially spread (blue curve), since the data is not aligned. The standard deviation is 4 cm (which coincides with the x-axis extents of the figure). After the first iteration (green curve), the curve already resembles a Gaussian distribution. While large progress is made during the first iterations, later iterations lead only to minor changes (cyan, magenta, yellow curves). The final standard deviations are 3.5 mm (2 cm threshold/ 897 M points remaining), 2.8 mm (1 cm/ 873 M) and 2.5 mm (7 mm/ 849 M).

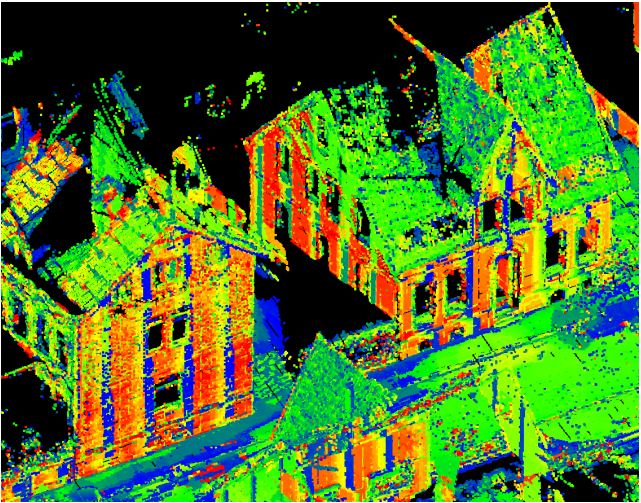

**Figure 15. Detail view, before estimation. Color scale: standard deviation, temperature scale, blue (0 mm) to red (100 mm or more).**

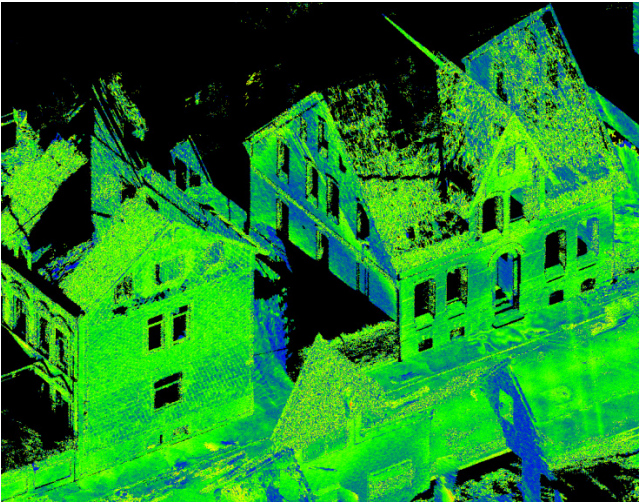

**Figure 16. Same view as Figure 15, after estimation. Note the color scale has been changed to the range between 0 mm (blue) to 7 mm (red). Compare the lower left building to Figure 20.**

## 7.2 Qualitative Results

The effect of the adjustment can be easily seen in the adjusted point clouds. Figure 18 shows the alignment of facades and walls in the scene (the colors indicate the original segmentation). Figure 19 shows a tilted view, where the increased sharpness is obvious, especially at the lattice fence in front. Note that for these visualizations, only the points of the point cloud are included which were assigned to a region during the pre-segmentation step. However, the results of the adjustment may also be applied to the original point cloud.

Figure 20 shows the latent map for the building in the lower left of Figure 16, with 1 cm raster size. Note how the surface structure of the roughcast in the building base and the different shingle tiling patterns in the first and second floor are visible. Figures 21 and 22 show two more examples where the precision and detail of the latent map become apparent.

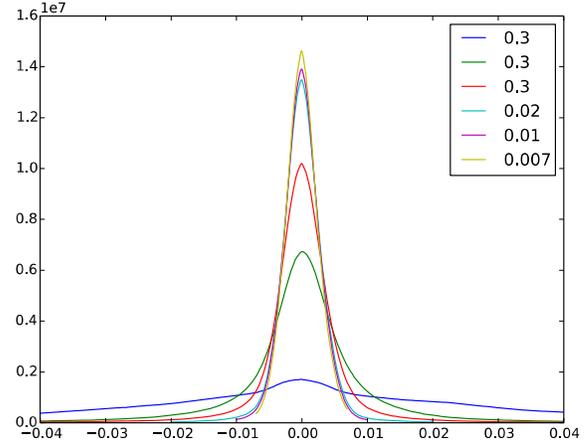

**Figure 17. Histograms of the signed distance of LiDAR points to the latent map. Initial distribution (blue), after first (green) and second (red) iteration (all with a distance threshold of 0.3 m), and after iterations with 2 cm (cyan), 1 cm (magenta) and 7 mm (yellow).**

## 8. CONCLUSIONS AND OUTLOOK

This paper presents a global strip adjustment method to obtain dense surface maps with a precision of a few millimeters for arbitrary large scenes. In the example shown here, it is applied to a scene of one billion points, estimating 278,000 unknowns.

The method uses a pre-processing MapReduce step for finding continuous regions and building up a representation in terms of spatial tiles. During the map phase, each scan strip is segmented individually, and the results are sent to the correct spatial cell during shuffle.

In the main part of the algorithm, an alternating least squares approach is used, which is realized as a number of MapReduce steps. In the map phase, each tile generates a set of normal equation blocks. During shuffling, these are grouped by trajectory. Finally, in each reducer, the maximum likelihood estimate is obtained for each trajectory individually using the highly efficient Rauch-Tung-Striebel algorithm.

The method scales linearly in time and space. In pre-processing, it is linear in the number of scan strips. During estimation, the map phase is linear in the number of tiles, while the reduce phase is linear in the total length of the trajectories. Note that the algorithm does not need matrices beyond size $6 \times 6$, and at no point in the algorithm data needs to be funneled through a single machine. Therefore, the algorithm lends itself well for a scaling to arbitrary scene sizes using MapReduce on a distributed file system.

For the future, it is planned to increase the robustness of the method. Currently, scan correspondences are ultimately based on proximity. This could be improved by a stronger role of the segmentation or by the introduction of an additional classification step. Also, the estimation of additional calibration parameters may improve the results. Furthermore, a pre-processing step for coarse alignment may be useful, in order to extend the approach to low cost mapping vehicles, or even to a crowd mapping approach involving normal vehicles equipped with LiDAR and/ or camera sensors.

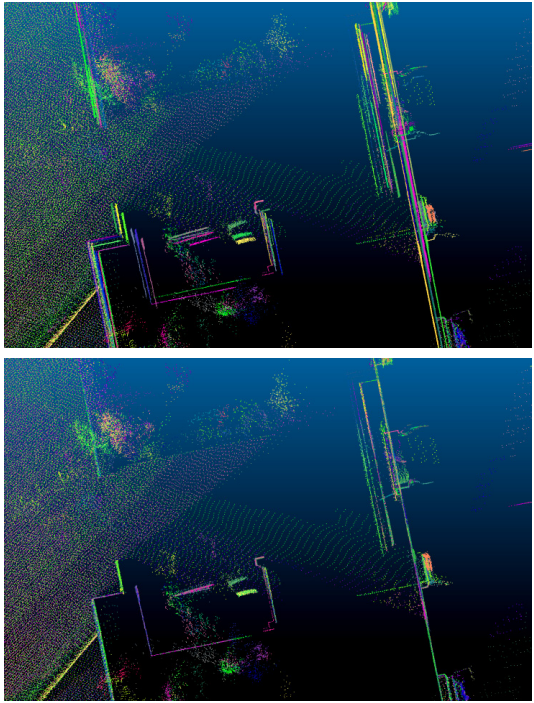

**Figure 18. Top view of a part of the point cloud, before (top) and after (bottom) adjustment. Note how the façade (right) and walls are well aligned after adjustment.**

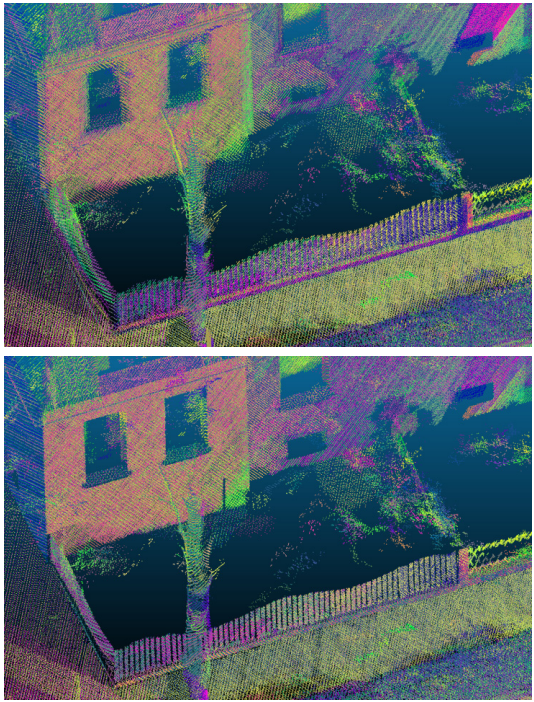

**Figure 19. Tilted view of a part of the point cloud, before (top) and after (bottom) adjustment. See especially the lattice fence.**

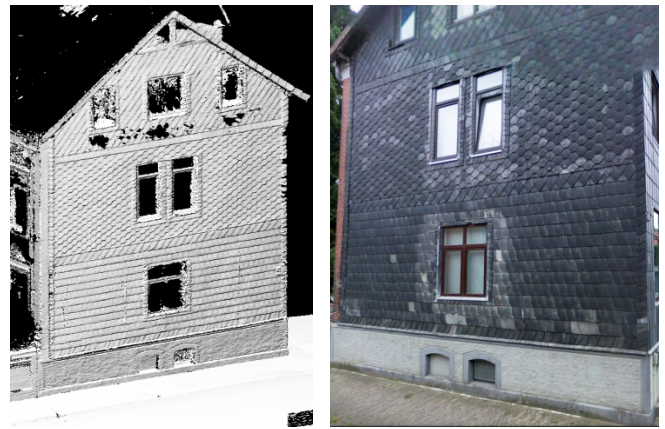

**Figure 20. Left: latent map for the lower left building in Figure 16. Note the details in the surface structure. Right: for comparison, StreetView image (image credit: Google).**

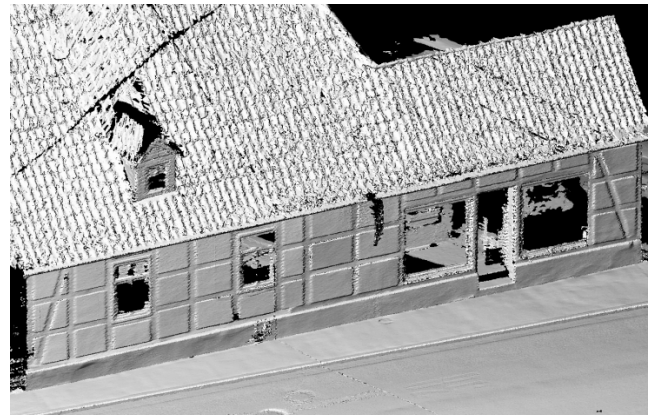

**Figure 21: Example of a half-timbered house where timber and protruding infill are clearly visible.**

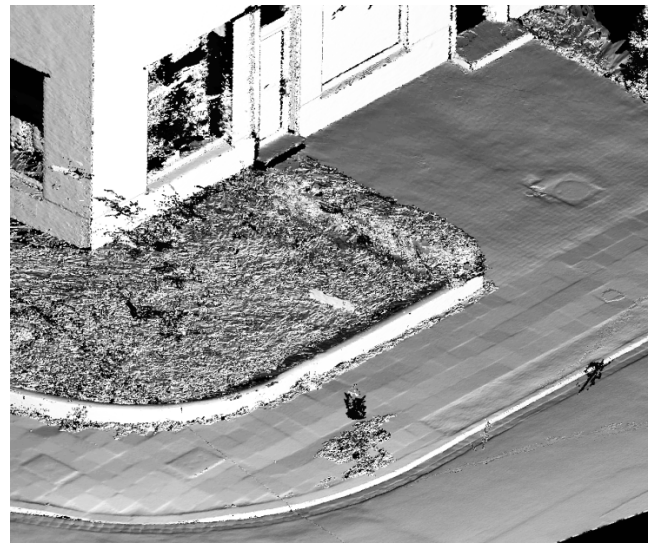

**Figure 22: Example with different surface structure. Flat tarred surface on the road, large tiles on the sidewalk and small tiles in front of the building.**

## 9. REFERENCES

[1] Applanix Corp. 2012. *POSLV Specifications.* http://www.applanix.com (June 23, 2016).

[2] Besl, P. J. and McKay, N. D. 1992. A method for registration of 3-D shapes. *IEEE Trans. Pattern Anal. Mach. Intell.* 14, 2, 239-256. DOI= http://dx.doi.org/10.1109/34.121791.

[3] Curless, B. and Levoy, M. 1996. A volumetric method for building complex models from range images. In *ACM Transactions on Graphics SIGGRAPH '96*, 303-312. DOI= http://dx.doi.org/10.1145/237170.237269.

[4] Dean, J., Ghemawat, S. 2004. MapReduce: Simplified Data Processing on Large Clusters. In *OSDI'04: Sixth Symposium on Operating System Design and Implementation* (San Francisco, CA, December, 2004).

[5] Felzenszwalb, P. and Huttenlocher, D. 2004. Efficient Graph-Based Image Segmentation. *International Journal of Computer Vision*, 59, 2, 167-181. DOI= http://dx.doi.org/10.1023/B:VISI.0000022288.19776.77.

[6] Glira, P., Pfeifer, N. and Mandlburger, G. 2015. Rigorous Strip adjustment of UAV-based laserscanning data including time-dependent correction of trajectory errors. In *9th International Symposium of Mobile Mapping Technology, MMT 2015* (Sydney, Australia, 9-11 December, 2015).

[7] Grisetti, G., Kümmerle, R., Stachniss, C. and Burgard, W. 2010. A Tutorial on Graph-Based SLAM. *IEEE Intelligent Transportation Systems Magazine* 2, 4, 31-43. DOI= http://dx.doi.org/10.1109/MITS.2010.939925.

[8] Huang, Q.-X. and Anguelov, D. 2010. High quality pose estimation by aligning multiple scans to a latent map. In *IEEE International Conference on Robotics and Automation (ICRA)* (Anchorage, AK, 2010), 1353-1360. DOI= http://dx.doi.org/10.1109/ROBOT.2010.5509460.

[9] Kilian, J., Haala, N. and Englich, M. 1996. Capture and Evaluation of Airborne Laser Scanner Data. *International Archives of Photogrammetry and Remote Sensing*, Vol. 31/3, ISPRS, Vienna, Austria, 383-388.

[10] Klingensmith, M., Dryanovski, I., Srinivasa, S., and Xiao, J. 2015. Chisel: Real Time Large Scale 3D Reconstruction Onboard a Mobile Device using Spatially Hashed Signed Distance Fields. In *Robotics: Science and Systems* (Rome, Italy). DOI= http://dx.doi.org/10.15607/RSS.2015.XI.040.

[11] Klingner, B., Martin, D. and Roseborough, J. 2013. Street View Motion-from-Structure-from-Motion. In *IEEE International Conference on Computer Vision* (Sydney, NSW, 2013), 953-960. DOI= http://dx.doi.org/10.1109/ICCV.2013.122.

[12] Koller, D., Friedman, N. 2009. *Probabilistic Graphical Models: Principles and Techniques*. MIT Press.

[13] Newcombe, R. A., Izadi, S. Hilliges, O., Molyneaux, D., Kim, D., Davison, A., Pushmeet, K., Shotton, J., Hodtes, S., and Fitzgibbon, A. 2011. KinectFusion: Real-time dense surface mapping and tracking. In *IEEE International Symposium on Mixed and Augmented Reality (ISMAR),* 127-136. DOI= http://dx.doi.org/10.1109/ISMAR.2011.6092378.

[14] Nießner, M., Zollhöfer, M., Izadi, S., and Stamminger, M. 2013. Real-time 3D reconstruction at scale using voxel hashing. *ACM Transactions on Graphics* (*SIGGRAPH Asia 2013*), 32, 6 (November 2013), 169:1-169:11. DOI= http://dx.doi.org/10.1145/2508363.2508374.

[15] Rauch, H. E., Tung, F., and Striebel, C. T., 1965. Maximum likelihood estimates of linear dynamic systems. *AIAA Journal*, 3, 8, 1445-1450. DOI= http://dx.doi.org/10.2514/3.3166.

[16] Riegl LMS GmbH 2012. *Data sheet Mobile Mapping System RIEGL VMX-250*. http://www.riegl.com (June 23, 2016).

[17] Rusinkiewicz, S. and Levoy, M. 2001. Efficient variants of the ICP algorithm. In *Third International Conference on 3-D Digital Imaging and Modeling* (Quebec City, Que., 2001), 145-152. DOI= http://dx.doi.org/10.1109/IM.2001.924423.

[18] Segal, A. V., Haehnel, D., and Thrun, S. 2009. Generalized-ICP. In *Robotics: Science and Systems* (Seattle, USA, June). DOI= http://dx.doi.org/10.15607/RSS.2009.V.021.

[19] Triggs, B., McLauchlan, P. F., Hartley, R. I., Fitzgibbon, and Andrew W. 2000. Bundle Adjustment – A Modern Synthesis. In *Vision Algorithms: Theory and Practice: International Workshop on Vision Algorithms* (Corfu, Greece, September 21-22, 1999 Proceedings), 298-372. DOI= http://dx.doi.org/10.1007/3-540-44480-7_21.